\documentclass[aps,prl,twocolumn,amsmath,amssymb,amsbsy,
superscriptaddress,floatfix,showpacs]{revtex4}
\usepackage[dvips]{graphicx}
\begin{document}
\title{Nematic order in square lattice frustrated ferromagnets}
\author{Nic~Shannon}
\affiliation{Max--Planck--Institut f{\"u}r Chemische Physik fester
Stoffe, N{\"o}thnitzer Str. 40, 01187 Dresden, Germany}
\affiliation{H. H. Wills Physics Laboratory, University of Bristol,
Tyndall Ave, BS8-1TL, UK}
\author{Tsutomu~Momoi}
\affiliation{Condensed Matter Theory Laboratory, RIKEN, Wako,
Saitama 351-0198, Japan}
\author{Philippe~Sindzingre}
\affiliation{
Laboratoire de Physique Th\'eorique de la Mati\`ere Condens\'ee, UMR
7600 of CNRS,  \protect\mbox{Universit\'e P. et M. Curie, case 121,
4 Place Jussieu, 75252 Paris Cedex, France}}
\date{\today}
\begin{abstract}
We present a new scenario for the breakdown of ferromagnetic order
in a two--dimensional quantum magnet with competing ferromagnetic
and antiferromagnetic interactions. In this, dynamical effects
lead to the formation of two--magnon bound states, which undergo
Bose--Einstein condensation, giving rise to bond--centered nematic
order. This scenario is explored in some detail for an extended
Heisenberg model on a square lattice.   In particular, we present
numerical evidence confirming the existence of a state with
d--wave nematic correlations but no long range magnetic order,
lying between the saturated ferromagnetic and collinear
antiferromagnetic phases of the ferromagnetic $J_1$--$J_2$ model.
We argue by continuity of spectra that this phase is also present
in a model with 4--spin cyclic exchange.
\end{abstract}
\maketitle

The search for a true quantum
``spin liquid''  --- a quantum magnet which remains disordered at
the very lowest temperatures ---  in dimension greater than one
has been central to research on quantum magnets for more than
three decades. Following Anderson \cite{rvb}, most models of
quantum spin liquids proposed to date have been based on
frustrated antiferromagnetic (AF) interactions. The resulting spin
liquid states involve strong
singlet bonds between spins, which give rise to a gap in the spin
excitation spectrum (for a review, see e.g.~\cite{ml03}).

None the less, the best characterized experimental realization of
a quantum spin liquid is believed to occur in two--dimensional
films of solid $^3$He, where the interactions between spins are
predominantly ferromagnetic (FM), and the resulting state is
gapless~\cite{Fukuyama,Ishimoto}. This raises the interesting
question of whether the breakdown of long ranged FM order offers a
new route to a spin liquid ground state~?

\begin{figure}[tb]
  \centering
  \includegraphics[width=7truecm]{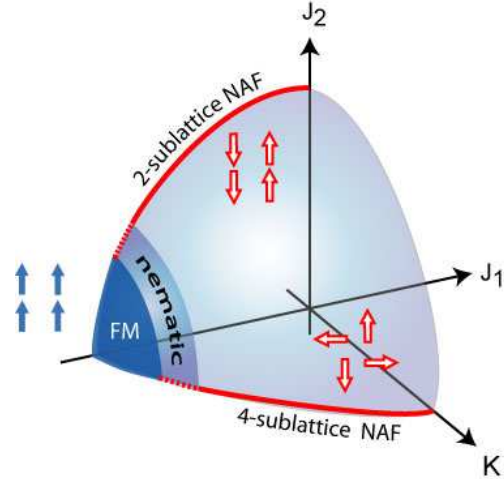}
  \caption{Phase diagram for the spin--1/2
  $J_1$--$J_2$--$K$
  Heisenberg model on a square lattice, in the octant where $J_1<0$ is
  FM and both $J_2\ge 0$ and $K \ge 0$ are AF.   Solid lines in the
  $J_1$--$J_2$ and $J_1$--$K$ planes denote known N\'eel
  phases.   A  single nematic phase interpolates between the FM
  occurring for large FM $J_1$ and these
  two, different, AF ordered states.
\label{fig1}}\end{figure}

The purpose of this paper is to demonstrate that a gapless spin
liquid can indeed occur in a simple two--dimensional model of
quantum spins on a square lattice with predominantly FM
interactions. To this end we present a set of analytic and numerical
results which show how a new type of ``bond--nematic'' state emerges
from the ruins of FM order in an extended Heisenberg model on a
square lattice. Our work is inspired by a number of new FM square
lattice compounds~\cite{dresden,kageyama} and has close parallels in
earlier work on spin chains~\cite{andrey}. The special case of the
multiple spin exchange model on a triangular lattice
--- which is believed to describe the magnetism of two--dimensional solid $^3$He
--- will be dealt with elsewhere~\cite{momoi}.

While these results are rather general, for concreteness in this
Letter we consider a square lattice $S=1/2$ frustrated ferromagnet
defined by the Hamiltonian
\begin{eqnarray}
    \label{eqn:H}
{\mathcal H} &=&
2 J_1 \sum_{\langle ij \rangle_1} {\bf S}_i \cdot {\bf S}_j + 2
J_2 \sum_{\langle ij \rangle_2} {\bf S}_i \cdot {\bf S}_j
\nonumber\\
&& + K \sum_{\langle 1234 \rangle} (P_{1234} + P_{1234}^{-1}) -h
\sum_i  S_i^z,
\end{eqnarray}
where $\langle ij \rangle_1$ counts nearest neighbor bonds, $\langle
ij \rangle_2$ counts next--nearest neighbor bonds, and $P_{1234}$
performs a cyclic exchange of spins on a single square plaquette. We
restrict ourselves to the case where $J_1 < 0$ is FM and the other
two interactions $J_2 \ge 0$ and $K \ge 0$ are AF.

\begin{figure}[tb]
  \centering
   \includegraphics[width=7truecm]{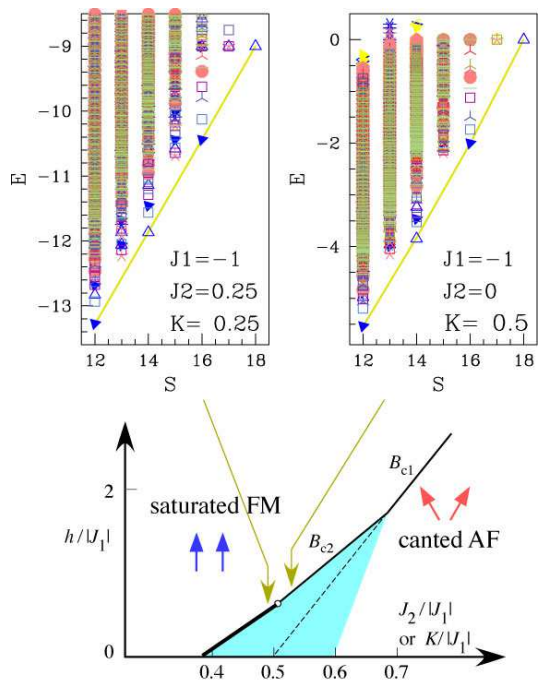}
  \caption{Schematic phase diagram of the FM
  $J_1$--$J_2$--$K$ extended Heisenberg model in applied field, showing
  how
  the two--magnon instability crosses over from 1$^{st}$ to 2$^{nd}$
  order in applied field and is eventually succeeded by a
  conventional one--magnon instability.   The inset shows
  representative spectra for 1$^{st}$ and 2$^{nd}$ order transitions
  into d--wave nematic state (high spin states only).
  \label{fig2}}
\end{figure}

The essence of our results, summarized in Fig.~\ref{fig1} and
Fig.~\ref{fig2}, is as follows
---  the competition between FM and
AF interactions leads to  a large ``accidental'' degeneracy  in
the one--magnon spectrum of the FM phase, which then becomes
highly susceptible to the formation of two--magnon bound states.
These condense, and give rise to a new singlet phase lying between
the classical FM and known N\'eel phases of the model. This phase
has a well defined Goldstone mode, but no long--range magnetic
order in the conventional sense.    We introduce an appropriate
order parameter for this phase, and identify it as a d--wave
n--type ``bond--nematic'' state, distinct from conventional
``n--type'' and ``p--type'' nematic states known to occur in other
closely related models \cite{harada,andreas}. From numerical exact
diagonalization calculations we are able both to confirm that this
phase exists, and to establish that it interpolates between the
two well--studied limits of Eq.~(\ref{eqn:H}), the square lattice
$J_1$--$J_2$ model for $K \to 0$ and the square lattice $J_1$--$K$
multiple spin exchange model for $J_2 \to 0$.

We now turn to the analysis.  For large FM $J_1$ the ground state of
Eq.~(\ref{eqn:H})  is a uniform saturated FM.     In the FM
$J_1$--$J_2$ model,  for $K=0$ this gives way to 2--sublattice
collinear N\'eel AF at large $J_2$ \cite{j1j2}.   In the FM
$J_1$--$K$ model,  for $J_2=0$, the corresponding N\'eel state is an
orthogonal 4--sublattice structure \cite{j1K}. Our analysis begins
with a very simple observation --- at a classical level, the
transition between FM and N\'eel states is 2$^{nd}$ order and takes
place where $|J_1| = 2(K + J_2)$.  However,  exactly at this
transition, the one---magnon spectrum of the FM phase posses entire
lines of zero modes \cite{nic2004}. This ``accidental'' degeneracy
has profound consequences for magnetism.

Firstly, it leads to large zero point fluctuations which shift the
first instability of the FM to somewhat weaker AF coupling.   We
have performed numerical exact diagonalization (ED) of
Eq.~(\ref{eqn:H}) for clusters of $16$, $20$, $32$ and $36$
spins~\cite{philippe}. The evolution of the ground state energy
indicates that a 1$^{st}$ order transition from saturated FM to
singlet ground states occurs for $|J_1| \approx 2.5(K + J_2)$.

Secondly, these fluctuations also destabilize the neighboring N\'eel
phase. Semi--classical estimates for the the square lattice
$J_1$--$J_2$ model suggest that the sublattice magnetization of its
collinear N\'eel phase vanishes for $|J_1| \approx  1.97
J_2$~\cite{nic2004}.
These calculations almost certainly underestimate quantum effects.

These arguments provide good {\it a priori} reason to believe that
the extended Heisenberg model Eq.~(\ref{eqn:H}) supports a new
spin liquid phase for a finite range of parameters between its FM
and N\'eel phases. However they give little insight into what this
phase might be.

In order to better answer this question, we introduce magnetic
field $h$ and examine the nature of the first instability of the
fully saturated paramagnet as the magnetic field is reduced. The
situation is summarized in Fig.~\ref{fig2}. For parameters well
within the classical N\'eel phases of Eq.~(\ref{eqn:H}), the
first instability of the saturated state occurs at $h = B_{c1}$
and is controlled by the lowest one--magnon excitation in applied
field. This instability is against a conventional two--sublattice canted
AF state.

In the proposed spin liquid region, however, the one--magnon
spectrum is highly degenerate.    In the extreme case, $J_2 = 0$, $K
= |J_1|/2$, the one--magnon dispersion vanishes altogether, so there
is no net energy gain in flipping one spin at any ${\bf q}$ and the
single flipped spins are always localized. On the other hand, {\it
pairs} of flipped spins on neighboring sites {\it can} propagate
coherently under the action of the cyclic exchange operator
$P_{1234}$, and so gain kinetic energy. It is therefore reasonable
to ask whether the transition out of the  FM phase controlled by
such two--magnon bound states ?

Both the one--body problem of a single flipped spin--1/2 in a
saturated FM background $|{\rm FM}\rangle$ and the two--body
problem of two interacting flipped spins can be solved exactly.
We have calculated the energy of both states in the
proposed spin liquid region and find that, at a critical value of field
$h=B_{c2}$, two--magnon bound states of the form
$    | \phi \rangle  = \sum_{ij} \phi_{ij} S^-_i S^-_j |{\rm FM}  \rangle$
with d-wave symmetry are gapless and give the first instability,
while one--magnon states have higher positive energy --- see
Fig.~\ref{fig2}. The binding energy of the two--magnon state is of
purely kinetic origin. In zero field, the two--magnon bound states
first become stable in the $J_1$--$J_2$ model for $J_2/|J_1| =
0.408$ (c.f.~\cite{dimitriev97}), and in the $J_1$--$K$ model for
$K/|J_1| = 0.364$. These values should be compared with the
numerically determined boundary of the FM phase at $(J_2 + K)/|J_1|
\approx 0.4$. In the special case $J_2 = 0$, $K = |J_1|/2$, the
two--magnon bound state has the compact form \mbox{$
    \phi_{ij} = \frac{1}{\sqrt{2}} [ \delta_{i,j-e_1} -
    \delta_{i,j-e_2}]
$}, where ${\bf e}_1 = (1, 0)$ and ${\bf e}_2 = (0, 1)$ connect
neighboring sites on a square lattice.

What then happens below $B_{c2}$ ?   In the case of the FM
$J_1$--$K$ model for $J_2$ = 0, the answer is quite simple.
Two--magnon pairs undergo Bose--Einstein condensation. For this set
of parameters there is a weak, repulsive interaction between
two--magnon pairs, so the transition is 2$^{nd}$ order. This is
evident from ED spectra --- the locus of the lowest lying state in
the high--spin sector has a {\it convex} curvature as a function of
$S$, so the magnetization of the system must evolve continuously at
$B_{c2}$ --- see the inset to Fig.~\ref{fig2}. Even spin sector
states only appear in the lowest states in magnetic field. This
period two is the evidence of condensation of two-magnon pairs.

In the FM $J_1$--$J_2$ model for $K=0$ the situation is a little
more complicated.   The net interaction between two--magnon pairs
is attractive, and the transition is 1$^{st}$ order.   Again this
is evident in ED from the  {\it concave} curvature of the high
spin states, which signals a jump in magnetization at saturation
field. A similar 1st order transition appears in the $J_1$--$K$
model for $J_2=0$ well close to the FM phase boundary. However in
all of these cases the structure of the low lying states is
essentially the same --- even spin sector states always have lower
energy than odd spin sector states, and show an alternation
between the s--wave $A_1$ and d--wave $B_1$ irreps of the square
lattice space group $C_{4\nu}$. We return these points below.

Where bound pairs of magnons condense they give rise to a new form
of bond--nematic order.   This is to say that, while the
orientation of individual spins remains undetermined, the
traceless rank two tensor
\begin{equation}
    \label{eqn:n-type}
{\cal O }^{\alpha\beta}(\boldsymbol{r}_i, \boldsymbol{r}_j)
= \frac{1}{2}(S^\alpha_i S^\beta_j+S^\beta_i S^\alpha_j) - {1
\over 3} \delta^{\alpha\beta}\langle \boldsymbol{S}_i \cdot
\boldsymbol{S}_j \rangle
\end{equation}
{\it does} exhibit long range order~\cite{shonan}.   The matrix elements of
${\cal O }^{\alpha\beta}(\boldsymbol{r}_i, \boldsymbol{r}_j)$ are
connected to the magnon pairing operator through the relation
$S_i^- S_j^- = {\cal O }^{xx} - {\cal O}^{yy} - 2i{\cal O }^{xy}$.
Physically, one can think of this tensor operator as revealing the
order ``hidden'' in the spin--1/2 quantum spin--liquid by projecting
into a symmetrized spin--1 Hilbert space of bond variables with long
range correlations.

The order parameter Eq.~(\ref{eqn:n-type}) is distinct from the
site--centred n--type nematic order seen in certain models with $S
= 1$ \cite{harada}, and from the p--type chiral order found in the
multiple spin exchange model with AF interactions \cite{andreas}.
However a very similar order parameter
was previously introduced in the context
of one--dimensional frustrated spin chains~\cite{andrey}.

\begin{figure}[tb]
  \centering
  \includegraphics[width=7truecm]{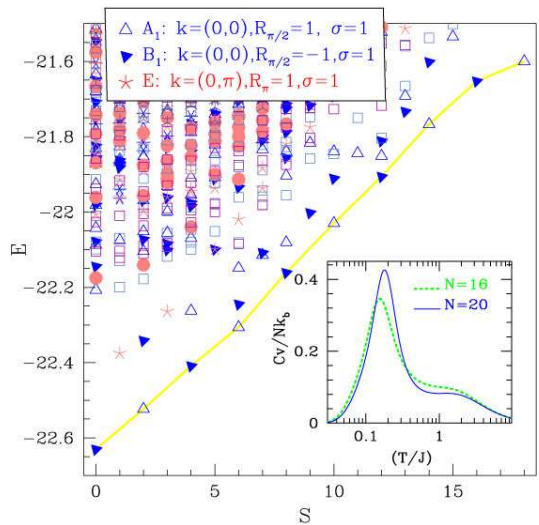}
  \caption{Spectrum of extended Heisenberg model for $N=36$ spin
  cluster with $J_1 = -1$,
  $J_2=0.4$, $K=0$, showing the marked alternation in energies of
  the lowest lying states in even and odd spin sectors.   The inset
  shows the double peak structure of the heat capacity $c_V(T)$
  for clusters of $N=16$ and $N=20$ spins and the same set of parameters.
\label{fig3}}
\end{figure}

Having established the unconventional nature of the state near
saturation, we now turn to the nature of the ground state in the
absence of applied field.   Proof that this is a spin liquid
follows from the finite--size scaling properties of ED spectra.
In Fig.~\ref{fig3} we present the spectrum of a cluster of
$N=36$ spins for $J_1 = -1$, $J_2 = 0.4$ and $K=0$, very close to
the boundary with the saturated FM phase.
The signature of long
range N\'eel order in such a spectrum would be the existence of a set of
``quasi--degenerate joint states'' (QDJS) which form the
$N=\infty$ ground-sate, presents in  every spin sector
(up to $S \sim \sqrt{N}$), with energies scaling as
\mbox{$E_{\rm QDJS} \sim \frac{S(S+1)}{N}$} and are
well separated (at finite $N$) from the lowest excitations
of the continuum (magnons with energies scaling as $\sim \frac{1}{\sqrt{N}}$).
The QDJS specific to $(\pi, 0)$ collinear AF order 
comprise one $A_1$ and one $B_1$ state for even $S$, and
a (twofold degenerate) $E$ level for $S$ odd~\cite{QDJS}.
The gaps to the low lying states in each spin sector should evolve as
\mbox{$\Delta \sim \frac{1}{N}[\alpha - \frac{\beta}{\sqrt{N}} ]$},
with coefficients $\alpha$ and $\beta$ determined by the spin
stiffness and spin wave velocity of the N\'eel state,
and thus diplay a negative curvature when plotted against $1/N$~\cite{gaps}.

\begin{figure}[tb]
  \centering
  \includegraphics[width=6truecm]{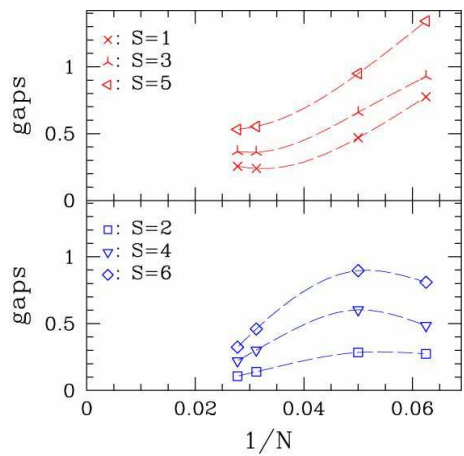}
  \caption{ Finite size scaling of the energy gaps of extended Heisenberg
  model with $J_1 = -1$, $J_2=0.4$, $K=0$, for $A_1$ and $B_1$ symmetry
  states (lower) and $E$ ones (upper) in the  even and odd spin
  sectors, respectively.
\label{fig4}}
\end{figure}

It is immediately clear from the spectrum in Fig.~\ref{fig3} that
the energies of the odd spin states, which include the $E$
symmetry states, are well separated from those of the even spin
states. Thus the spectrum is not compatible with the expected
N\'eel order. The unconventional nature of the ground state is
also reflected in the finite size scaling of gaps, shown in
Fig.~\ref{fig4}. The gaps in the even spin sectors scale to zero
in the expected manner. But the gaps of $E$ symmetry states in odd
spin sectors do not
--- in fact they appear to scale to a finite value $\Delta_{\rm
odd} \sim 0.4$.

This is conclusive proof that the FM $J_1$--$J_2$ model {\it does
not} exhibit N\'eel order in the immediate neighborhood of its FM
phase. But we can draw a still stronger conclusion. The spin
even-odd oscillation found in high spin states persists down to low
spin states, and the $A_1$ and $B_1$ states found in even spin
sectors are precisely the same states which emerge from the two
magnon instability of the saturated FM in applied magnetic field.
The existence of this set of QDJS is the signature of nematic order
with d--wave symmetry. Thus the nematic order found near to
saturation persists down to zero magnetic field.

\begin{figure}[tb]
  \centering
  \includegraphics[width=7truecm]{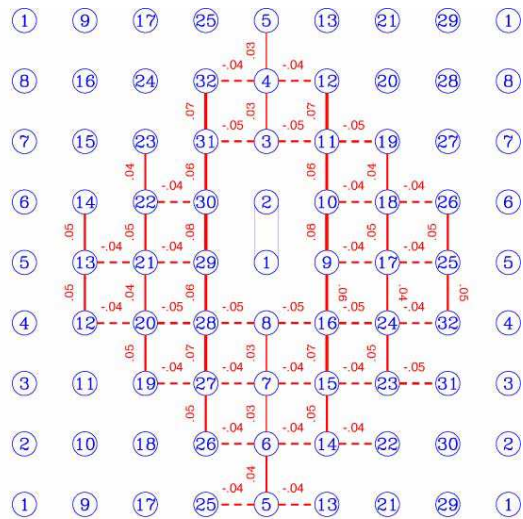}
  \caption{Bond--nematic correlations of an extended Heisenberg
  model with $J_1 = -1$, $J_2=0.4$, $K=0$, for a cluster of $N=32$ spins,
  showing the d--wave character of the nematic phase.
  Correlations are measured relative to the reference bond $1$--$2$.
\label{fig5}}
\end{figure}

Such nematic order has a degenerate ground--state composed of even
spin states. It has gapless Goldstone modes analogous to the lowest
magnons states of the N\'eel state, present in the $S=1$ sector,
scaling as $\frac{1}{\sqrt{N}}$. Results for a prototypical nematic
correlation function $\sum_{\alpha,\beta}\langle {\cal
O}^{\alpha\beta}({\bf 0},{\bf e}_1) {\cal O}^{\alpha\beta}({\bf
r},{\bf r}+{\bf e}_a)\rangle$ ($a=1,2$) in the nematic ground state
are presented in Fig.~\ref{fig5}. Nematic correlations exhibit a
``stripey'' character, and exist in the whole system. Because of the
underlying d--wave symmetry, correlations on parallel bond have
positive sign, while correlations on perpendicular bonds are
negative.

The finite temperature properties of the FM $J_1$--$J_2$ model
provides further insight into the bond--nematic state.
In an inset to Fig.~\ref{fig3}
we show the heat capacity calculated for the same exchange
parameters.   It exhibits two clear energy scales --- a broad structure
extending to $T \sim |J_1|$ which reflects the formation of $S=1$
objects from the paramagnetic ``soup'', and sharp peak at $T \sim
0.1 |J_1|$, where these $S=1$ objects start to order nematically.

Although it is hard to put a precise boundary on the stability of
the $(\pi, 0)$ collinear AF state, our ED studies suggest that
N\'eel order begins to break down in the FM $J_1$--$J_2$ model for
$J_2 \sim 0.6$--$0.7 |J_1|$.   We also have performed ED
calculations for parameter sets interpolating between the FM
$J_1$--$J_2$ model for $K=0$ and the FM multiple spin exchange
model for $J_2=0$. These confirm that the nature of the state
bordering the FM phase does not change.    Thus a single nematic
phase interpolates between the two different N\'eel order
parameters in these two limits, and the FM phase for large
$|J_1|$, as shown in Fig.~\ref{fig1}.

To conclude --- in the frustrated square lattice FM's 
considered in this letter, the saturated FM ground state undergoes a
first order transition into a gapless spin--liquid state with
bond--nematic order.  In applied magnetic field, the transition
becomes second order and can be related directly to
the condensation of two--magnon bound states.

It is our pleasure to acknowledge stimulating discussions with
Andreas L\"auchli, Claire Lhuillier and Karlo Penc. Computations
were performed at IDRIS (Orsay). This work was supported by
Grant-in-Aid for Scientific Research 
from MEXT of Japan and SFB 463 of the DFG.

\end{document}